\documentclass[11pt]{article}
\usepackage{moriond,epsfig}

\def\be{\begin{equation}}
\def\ee{\end{equation}}
\def\bea{\begin{eqnarray}}
\def\eea{\end{eqnarray}}

\usepackage{color}
\usepackage{graphicx}
\definecolor{red}{rgb}{1,0,0}
\def\lesssim{\ \hbox{\raise 2pt \hbox{$<$} \kern -13pt
                     \lower 3pt \hbox{$\sim$}}\ }
\def\greatersim{\ \hbox{\raise 2pt \hbox{$>$} \kern -13pt
                     \lower 3pt \hbox{$\sim$}}\ }

\def\herwig{{\sc Herwig}}
\def\pythia{{\sc Pythia}}

\input epsf.tex
\def\desepsf(#1 width #2){\epsfxsize=#2 \epsfbox{#1}}

\usepackage{amsmath,bm}

\begin{document}

\title{JETS IN  THE FORWARD REGION AT THE LHC~\footnote{Presented at the 
XLIV Rencontres de Moriond,  {\em QCD and High Energy  Interactions},
  March 2009.}}

\author{M.~Deak$^1$, F.~Hautmann$^2$, H.~Jung$^1$ and K.~Kutak$^1$  }

\address{$^1$Deutsches Elektronen Synchrotron, D-22603 Hamburg\\ 
$^2$Theoretical Physics,  University of Oxford,  Oxford OX1 3NP}

\maketitle\abstracts{We  
discuss   jet   production     at  forward rapidities at the LHC.   
In this  region    QCD  logarithmic  corrections in the 
hard transverse momentum  and   
in the large rapidity  interval  may both  be 
 quantitatively   significant.  We describe results of 
 using  high-energy factorization techniques, which    
 allow one to take  into account 
 both kinds of   corrections to higher  orders in QCD.}

Experiments at the  Large Hadron Collider  (LHC) 
  will  explore  the forward region in high-energy  hadronic  collisions  
 both with   general-purpose 
 detectors and with dedicated instrumentation, including 
  forward  calorimeters   and proton 
  taggers.~\cite{cmsfwd,atlasfwd,fp420,cmstotem,grothe,heralhc}    
The physics   program in the forward region 
involves   a wide range of topics, from  
new particle discovery  processes~\cite{fp420,fwdhiggs,fwdmssm}   
to  new  aspects of   
strong interaction physics~\cite{heralhc,denterria}  to 
heavy-ion  collisions.~\cite{accardi03,heavy-ion-cmsnote}
Owing to  the  large center-of-mass energy    and 
  the good  experimental coverage at large rapidities,  
 it becomes possible for the first time to investigate    forward-region physics  
 with  high-$p_\perp$ probes.

\begin{figure}[htb]
\vspace{45mm}
\includegraphics{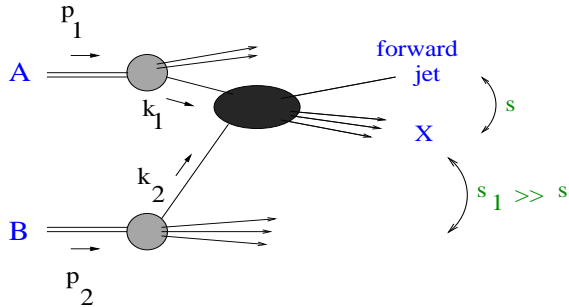}
\caption{Jet  production in the forward rapidity region 
in  hadron-hadron collisions.} 
\label{fig:forwpicture}
\end{figure}

 The hadroproduction of a  forward jet   associated  
with  hard final state $X$  is pictured 
in Fig.~\ref{fig:forwpicture}.    
The kinematics 
  of the process     is  characterized 
by the  large  ratio  of sub-energies  $s_1  /  s  \gg 1 $   
 and  highly asymmetric longitudinal momenta in the partonic initial 
  state, $k_1 \cdot p_2 \gg k_2 \cdot p_1$.  
 At the LHC the use of  forward calorimeters  allows  one to  
  measure    events  where   jet transverse momenta 
  $p_\perp  >   20$ GeV   are produced  several units of rapidity 
  apart,  $\Delta y   \greatersim 4 \div 6$.~\cite{cmsfwd,heralhc,aslano} 
 Working at    polar angles that are  small   but   sufficiently  far  from the beam axis 
 not to be affected by  beam remnants,     one measures 
 azimuthal plane correlations  
  between   high-$p_\perp$  events 
  widely  separated    in rapidity.~\cite{heralhc,preprint}

The   presence of multiple   large-momentum scales  
implies   that, as      recognized in~\cite{muenav,vddtang,stirl94},    
reliable  theoretical predictions   for forward jets 
can only be obtained after  summing  
 logarithmic  QCD corrections at high energy   
 to all orders in $\alpha_s$.  Analogous observation applies to  
  forward jets  associated 
 to  deeply    inelastic  scattering.~\cite{mueproc90c,forwdis92}   Indeed, measurements of 
 forward jet cross sections at Hera~\cite{heraforw,heraforw1} have illustrated that 
 either fixed-order next-to-leading  
 calculations  or  standard shower Monte Carlos, e.g.   \pythia\    or 
 \herwig,  are not  able to  
 describe 
 forward jet  $ep$ data.~\cite{heraforw1,web95,webetal99}     
This    motivates    efforts~\cite{webetal99,orrsti,stirvdd,andsab}  to 
 construct   new    
  algorithms  for   Monte Carlo  event  generators capable of 
   describing   jet   production    beyond  the central rapidity region. 
 
 It is   emphasized in~\cite{preprint}   that   
 in the  LHC forward   kinematics 
  realistic  phenomenology  of      hadronic  jet final states 
  requires taking into  account  at higher order   both     
  logarithmic corrections in    the large  rapidity  interval  
(of  high-energy type)  
and   logarithmic corrections 
 in  the hard transverse momentum (of collinear type).  
The theoretical framework to resum   consistently  
both  kinds of logarithmic corrections  in QCD 
 perturbation theory    is based on  high-energy    factorization at 
fixed transverse momentum.~\cite{hef}

\begin{figure}[htb]
\vspace{45mm}
\includegraphics{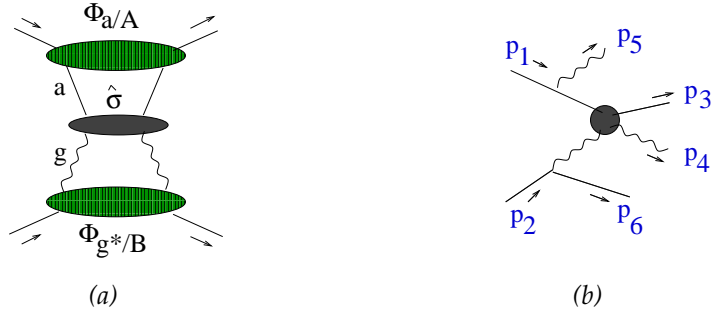}
\caption{(a) Factorized structure of the cross section; (b) a  graph  contributing   to the 
$q g$ channel matrix element.} 
\label{fig:sec2}
\end{figure}

The   high-energy  factorized    form  
 of the  forward-jet  cross section is   represented   in Fig.~\ref{fig:sec2}. 
Initial-state parton configurations  contributing to  
forward production are asymmetric, 
with the parton in the top subgraph being  probed near  the mass shell and  
large  $ x $,  
while  the parton in  the bottom subgraph is off-shell 
and small-$x$.\footnote{Studies  of   subleading corrections to jet production 
in the high-energy limit  are  given  in~\cite{schw0703},    with a view to 
applications to    measurements of  jet correlations.~\cite{mn_pheno,hj_ang}} 
The    jet  cross  section differential 
in the final-state   
transverse  momentum 
 $Q_T$  and  azimuthal angle $\varphi$ 
is given    by~\cite{preprint,hef}  
\begin{equation}
\label{forwsigma}
   {{d   \sigma  } \over 
{ d Q_T^2 d \varphi}} =  \sum_a  \int  \     
d \xi_1 \ d \xi_2  \  d^2 k_T \ 
 \phi_{a/A} (\xi_1)  \    {{d   {\widehat  \sigma}   } \over 
{ d Q_T^2 d \varphi  }}    ( \xi_1 \xi_2 S ,   k_T , Q_T , \varphi )  \ 
\phi_{g^*/B}  (\xi_2,   k_T)     \;\; ,
\end{equation}
where 
the sum goes over  parton  species, $\phi$ are  parton distributions  
defined from   unintegrated 
Green's functions,~\cite{mc98,ch94}  and 
$ {\widehat  \sigma} $  is the  hard-scattering  kernel,  calculable 
from  the high-energy limit  of  perturbative 
amplitudes  (Fig.~\ref{fig:sec2}b).   Results for  the  factorizing high-energy 
amplitudes are given in~\cite{preprint} in fully exclusive form. 
For phenomenological studies  it    is   of interest  to couple  
Eq.~(\ref{forwsigma}) to parton showers to achieve a full description   
of the associated final states.

 Work  to develop 
 methods for  the  evolution   of  unintegrated  parton  distributions 
 in parton-shower form 
  is currently 
   underway  by  several authors. 
 See for instance the   recent 
proposal,~\cite{jadach09}  which   incorporates 
NLO  corrections to    flavor non-singlet   QCD evolution in  
a shower Monte Carlo  at unintegrated level. 
The approach is based on the generalized ladder expansion of~\cite{CFP}. 
This  expansion 
is   extended to the  high-energy   
region in~\cite{ch94} and  used    to  resum    
 high-energy logarithmic corrections 
 to  space-like jet  evolution.   It   is  thus   likely that 
  the approach~\cite{jadach09}   can  
be applied more  
generally, including flavor-singlet evolution,     to   treat forward hard processes.   
 Overviews   of  theoretical  issues on unintegrated distributions   
and related   calculational  programs  can  be found 
in~\cite{jcc-lc08,hj_rec}.  

\begin{figure}[htb]
\vspace{50mm}
\includegraphics{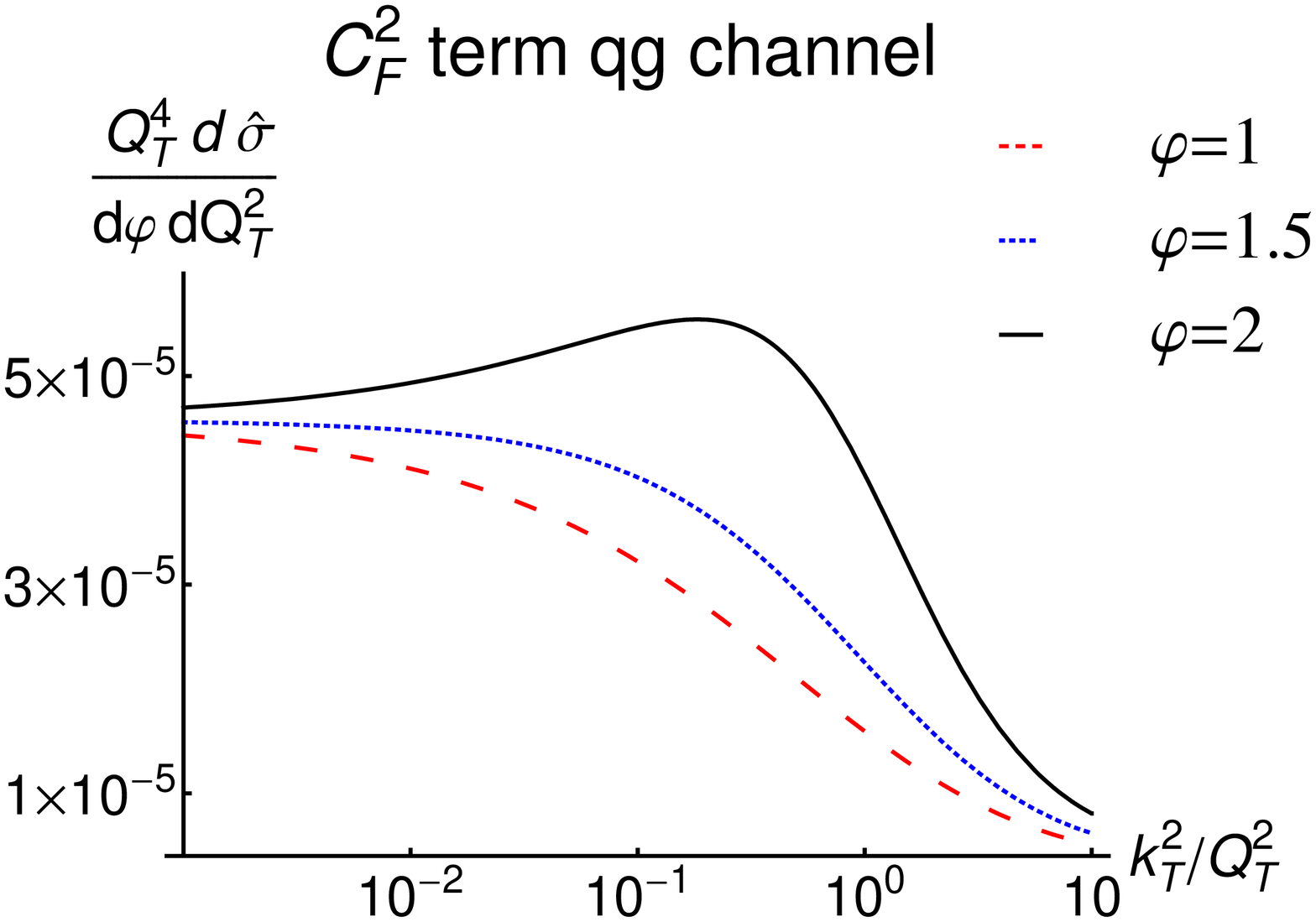}
\includegraphics{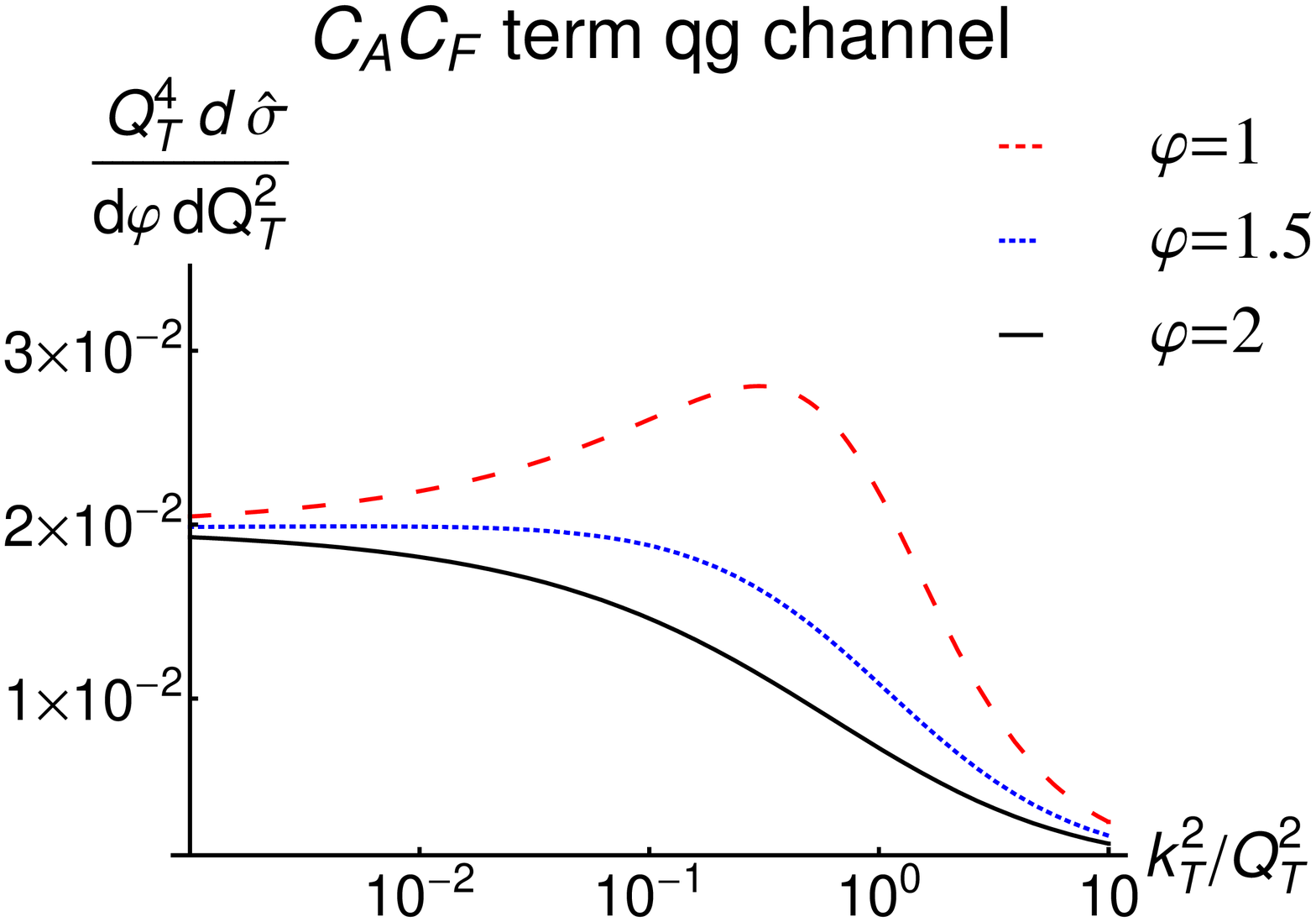}
\caption{The factorizing   $q g$   hard  cross 
section at high energy:$^{13}$          
 (left) $C_F^2$ term;  (right) $C_A   
C_F$ term.} 
\label{fig:forwplot}
\end{figure}

 Eq.~(\ref{forwsigma})   incorporates high-energy  corrections to forward jet production 
 through the  k$_\perp$ dependence of   both $\phi$ and  $ {\widehat  \sigma} $. 
 The  quantitative importance  of finite-k$_\perp$ corrections  
  is    associated with effects of coherence of 
 multiple gluon emission for small parton momentum fractions.~\cite{hef,mc98} 
  Potentially significant   coherence effects 
involve   both    the short-distance factor $ {\widehat  \sigma} $ and   the 
long-distance factor  $\phi$.  
An illustration is  shown   in Fig.~\ref{fig:forwplot},~\cite{preprint}   giving 
results  for the short-distance matrix element in the quark channel 
 $a = q$. 
We show separately   contributions  
  proportional to  color factors  $C_F^2$ and $C_A C_F$.   
In the notation of  Fig.~\ref{fig:sec2}, 
the final state transverse variable  $Q_T$ is defined as~\cite{preprint}  
\begin{equation}
\label{qtdef}
Q_T = (1-\nu) p_{T  4} - \nu p_{T  3}  \;\;, \;\;\;    {\rm{where}}  \;\;\; 
\nu =  (p_2 \, p_4) / [(p_2 \, p_1) -  (p_2 \, p_5)]   \;\;,  
\end{equation}
and  the azimuthal angle  
 $\varphi$ is measured with respect to the   di-jet transverse 
momentum.   
The curves in Fig.~\ref{fig:forwplot}   measure   
  the  $k_T$ distribution  
  of   the      jet  system    recoiling against the leading di-jets.
The leading-order   process with two back-to-back 
jets   corresponds to the region    $ k_T / Q_T \to 0$. 
 The dependence on   $ k_T   $ and $\varphi$   plotted 
in Fig.~\ref{fig:forwplot}     is the result of 
higher-order gluon radiation,   treated according 
to  the high-energy asymptotics.

  Fig.~\ref{fig:forwplot}  illustrates 
  that  the role of coherence  from  multi-gluon 
  emission is to set the dynamical cut-off   at values of 
  $ k_T $ of order  $ Q_T $.~\cite{preprint}   Non-negligible 
  effects arise at high energy    from the finite $k_T $  tail. 
These  effects are not included in  collinear-branching  
 generators  (and only partially in fixed-order 
 perturbative calculations),   
 and  become  more and more important  as  the jets are  observed at 
large rapidity separations.  
Monte Carlo implementations of    Eq.~(\ref{forwsigma})  and   
  the   coherent  matrix  elements     (Fig.~\ref{fig:forwplot})   
   are underway.~\cite{prepar} 
The usefulness of these  matrix elements comes from the fact 
 that  in the high-energy limit 
they factorize not only in the collinear emission region but also in the large-angle 
emission region.   It will be of interest to investigate  their role 
also with respect to  
the transverse-momentum and rapidity cut-off  scales~\cite{jcc-lc08,fhfeb07}  
that enter calculations in the  parton-shower  formalism at the unintegrated level.

A qualitatively similar behavior    to that   in    Fig.~\ref{fig:forwplot}  
  is observed in gluonic channels.~\cite{preprint}  
  We find   that 
  quark 
  and gluon  channels give contributions   of comparable size  to forward  jets 
 in  the    LHC    kinematics.   
Note also that 
since the forward kinematics selects 
asymmetric parton momentum fractions,    effects   
  due to    the   $ x \to 1$   endpoint   behavior~\cite{fhfeb07}   
   at  fixed transverse momentum  
may    become   phenomenologically   significant as well.

Let us  finally  recall   that  if effects  of  high-density parton 
matter~\cite{denterria}  
show up at  the LHC, they will affect 
high-$p_\perp$ forward physics.~\cite{heralhc,kugeratski}  
The  theoretical framework described above 
implies partonic  distributions dependent on both 
longitudinal and transverse degrees of freedom, and   
is likely   most natural to discuss issues of  parton saturation.  
Studies of forward jets in this context  
  are warranted.

{\bf Acknowledgments}.   Many thanks to the Moriond team  for the    
 invitation   and for    the excellent organization of the conference.

\end{document}